\documentclass{ws-ijmpd}

\begin{document}

\markboth{D. Ma and P. He} {Distribution Function of Dark Matter}

%
\catchline{}{}{}{}{}
%

\title{Distribution Function in Center of Dark Matter Halo}

\author{Ding Ma\footnote{Email: mading@itp.ac.cn}}

\author{Ping He\footnote{Email: hep@itp.ac.cn}}

\address{Institute of Theoretical Physics, \\
Chinese Academy of Sciences, P. O. Box 2735,\\
 Beijing 100190,
China}

\maketitle

\begin{history}
\received{Day Month Year} \revised{Day Month Year} \comby{Managing
Editor}
\end{history}

\begin{abstract}
N-body simulations of dark matter halos show that the density
profiles of halos behave as $\rho(r)\propto r^{-\alpha(r)}$, where
the density logarithmic slope $\alpha \simeq 1\sim1.5$ in the center
and $\alpha \simeq 3\sim 4$ in the outer parts of halos. However,
some observations are not in agreement with simulations in the very
central region of halos. The simulations also show that velocity
dispersion anisotropy parameter $\beta\approx 0$ in the inner part
of the halo and the so called "pseudo phase-space density"
$\rho/\sigma^3$ behaves as a power-law in radius $r$. With these
results in mind, we study the distribution function and the pseudo
phase-space density $\rho/\sigma^3$ of the center of dark matter
halos and find that they are closely-related.

\end{abstract}


\section{Introduction}\label{S1}
The formation and evolution of the dark matter halo which can be
treated as the self-gravitational collisionless stellar system have
become a challenging issue in the study of dark matter.

Thanks to the improved computational power, N-body simulations of
dark matter halos become more and more accurate and important with
increasing resolution. N-body simulations such as the universal NFW
profile\cite{NFW95,NFW96} and others\cite{M99,J2000,AR99} show that
the density profiles of dark matter halos behave as
$r^{-\alpha(r)}$, where $\alpha \simeq 1\sim1.5$ in the center and
$\alpha \simeq 3\sim 4$ in the outer parts of halos. However,the
numerical inner behaviors of dark matter halos are not supported by
observations.\cite{R85,Cour97,PW2000,Blok01,Blok03,Sal01,Sw03,Cor03}
Some work indicates that density profiles of dark matter halos might
become shallower than $r\simeq1$ in the center\cite{Austin05}, and
perhaps even tend to be a core with no cusp at
all.\cite{0608376,0603051}

N-body simulations not only provide us the density profiles of
halos, but also give the relevant information of the velocity space
of collisionless particles in the halos. Velocity dispersion and
anisotropy
profiles\cite{Rasia04,Mc05,Merritt06,Cole96,Carl97,Col00,FM01} have
been well described by simple analytical fits. Two interesting
phenomena from simulations indicate that velocity dispersion and the
density profiles of the haloes are not independent. First, Hansen
and Moore\cite{HM06,HS06} found that the density logarithmic slope
$\alpha(r)$ is correlated to the velocity anisotropy which is
parameterized by the anisotropy parameter $\beta(r)$ and they
provided the empirical formula $\beta\approx1-1.15(1-\alpha/6)$.
Therefore $\beta\approx 0$ (isotropic velocity dispersion) in the
inner part as $\alpha\approx 1$ and $\beta\approx 0.5$ in the outer
part as $\alpha\approx 3$. Second, it has been argued\cite{TN01}
that the so called pseudo phase-space density follows a power law
$\rho(r)/\sigma^3(r)\propto r^{-\gamma}$ with exponent
$\gamma=1.875$, where $\rho(r)$ is the density profile and
$\sigma^2$ is the total velocity dispersion. Subsequent studies have
confirmed that $\rho(r)/\sigma^3(r)$ is a power law in radius, but
the best fitting values of the exponent $\gamma$ diverse from each
other\cite{Rasia04,As04,Hof07,Mac06} and range from $\gamma=
1.90\pm0.05$ to $2.19\pm0.03$. Because $\rho/\sigma^3$ has the same
dimension as the phase-space density, $\rho/\sigma^3$ has been
called pseudo phase-space density or "poor-man's" phase-space
density\cite{0510332}.

With these results in mind, much theoretical work has been done for
the study of the relation between density profile behavior and
pseudo phase-space density. Some authors examine this matter by
solving the Jeans equation. Williams et al got a critical
exponents\cite{0412442} $\gamma=35/18$ and Dehnen and McLaughlin
calculated corresponding density profiles in both isotropic and
anisotropic cases\cite{DM05}. Some other work solved the Jeans
equation to explore the relation between density profile and pseudo
phase-space density \cite{Austin05,0510332,0405371,0609784}.
Recently, R. N. Henriksen\cite{0709.0434} considered a series
expansion for a dark matter distribution function in the spherically
symmetric anisotropic limit to discuss pseudo phase-space density.

In this paper, we concentrate on the center of the dark matter halo
where the velocity dispersion is almost isotropic and calculate the
distribution function and pseudo phase-space density where
$r\rightarrow 0$ in the spherically symmetric case. In Section 2 we
review the basic knowledge about distribution function which is
needed for this paper. In Section 3, the distribution function,
velocity dispersion and pseudo phase-space density in the center of
the dark matter halo are calculated with the given density profile.
We make the discussion and conclusion in Section 4.

\section{General Formulae}
In the center of dark matter halos, with spherically symmetric
assumption, the distribution function $F(E,L)$ can be reduced to
$F(E)$ as anisotropy parameter $\beta\approx 0$ (almost isotropic)
when $r\rightarrow 0$. If we know the distribution function, we can
calculate the density profile and total velocity dispersion profile
as below:\cite{De86},
\begin{equation}
 \rho(\psi)=4\pi \int_0^{\psi}F(E)\sqrt{2(\psi-E)}dE\ ,
\label{rhoiso}
\end{equation}

\begin{equation}
\sigma^2=\sigma^2_r+\sigma^2_T=\tfrac{2^{7/2}\pi
M_{tot}}{\rho}\,\int_0^{\psi}F(E)(\psi - E)^{3/2}dE, \label{sigma}
\end{equation}
where the binding energy $E$ is defined as
\begin{equation}
E=\psi(r) - \tfrac{1}{2}\,v_r^2 - \tfrac{1}{2}\,v_T^2\ , \label{E}
\end{equation}
and $\psi(r)$ is the relative gravitational potential which can be
obtained from the Poisson equation:
\begin{equation}
  \frac{1}{r^2}\,\frac{d}{dr}
  \left(r^2\,\frac{d\psi}{dr}\right)
  =-4\pi G\rho(r)\ .
  \label{eq:poisson}
\end{equation}
$v_r$ is the radial velocity and $v_T$ is the tangential velocity:
\begin{equation}
  v_T=\sqrt{v_\theta^2+v_\varphi^2}\ .
\end{equation}
Eddington\cite{Ed16} provided the inversion formula of
Eq.~(\ref{rhoiso}):
\begin{equation}
F(E)=\frac{1}{2\pi^2M_{tot}}D_E^2\int_0^E\frac{\rho(\psi)d\psi}
{\sqrt{2(E-\psi)}}\ , \label{Fiso}
\end{equation}
where $D_E^2$ denotes the 2nd order differentiation operator with
respect to $E$.

The anisotropy parameter\cite{Bi80} mentioned in the Introduction is
defined as:
\begin{equation}
\beta=1-\frac{\sigma_T^2}{2 \sigma_r^2}\ ,
\end{equation}
where $\sigma_T^2$ and $\sigma_r^2$ are the tangential and radial
velocity dispersion. If $\beta<0$, the velocity dispersion is
tangentially anisotropic. $0<\beta\leq1$ correspond to the radially
anisotropic cases, and $\beta=0$, the isotropic case.

\section{Distribution Function \& Pseudo Phase-Space Density}
In this Section, we will calculate the distribution function and
pseudo phase-space density with the given density profile in the
very central region of the spherically symmetric dark matter halo
where the velocity dispersion is almost isotropic. Let us consider a
family of density profiles with parameters $a$ and $\epsilon$ named
as New generalized NFW profiles,\cite{0803.1431}
\begin{equation}
\rho=C\frac{1}{(r/r_s)^a(1+r/r_s)^{3+\epsilon-a}}
 \label{eq:rho}\ .
\end{equation}

We set the characteristic radius $r_s=1$, the total mass $M_{tot}=1$
and the gravitational constant $G=1$ here. Then the density profiles
reduce to

\begin{equation}
\rho=C\frac{1}{r^a(1+r)^{3+\epsilon-a}}
 \label{eq:rho}\ ,
\end{equation}

\begin{equation}
C=\frac{\Gamma(3-a+\epsilon)}{4\pi\Gamma(3-a)\Gamma(\epsilon)},\
a<3, \epsilon>0
\end{equation}

Then the relative potential $\psi(r)$ can be calculated from the
Poisson Eq.~(\ref{eq:poisson}).\cite{0803.1431}
\begin{gather}
\begin{split}&
 \psi(r)=\psi_0 - 4\pi C
 r^{-a}\Gamma(-a)\left(A_1(r)-2A_2(r)+A_3(r)\right)\
\\&
\\&
 A_1(r)=\ _2\tilde{F}_1[1-a+\epsilon,\ -a,\ 2-a,\ -r]
\\&
 A_2(r)=\ _2\tilde{F}_1[2-a+\epsilon,\ -a,\ 2-a,\ -r]
\\&
 A_3(r)=\ _2\tilde{F}_1[3-a+\epsilon,\ -a,\ 2-a,\ -r]
 \label{eq:psi}
\end{split}
\end{gather}
where $\psi_0=\frac{\epsilon}{2-a}$ is the relative gravitational
potential at $r=0$ which is determined by the condition
$\left.\psi(r)\right|_{r\rightarrow\infty}=0$. And then, we
calculate the asymptotic approximation of
$\left.\psi(r)\right|_{r\rightarrow 0}$ and
$\left.\rho(\psi)\right|_{r\rightarrow 0}$ as below:

\begin{equation}
\left.\psi(r)\right|_{r\rightarrow 0}=\psi_0-\frac{4\pi C
}{6-5a+a^2}r^{2-a}
 \label{limitpsir}
\end{equation}

\begin{gather}
\begin{split}&
\left.\rho(\psi)\right|_{r\rightarrow
0}=B(\psi_0-\psi)^{-\frac{a}{2-a}}
\\&
\ \ B=C(\frac{6-5a+a^2}{4\pi C})^{-\frac{a}{2-a}}
 \label{rhopsi}
\end{split}
\end{gather}

From $\left.\psi(r)\right|_{r\rightarrow 0}$ and
$\left.\rho(\psi)\right|_{r\rightarrow 0}$, we can get the
asymptotic approximation of $\left.F(E)\right|_{r\rightarrow 0}$
when using Eq.~(\ref{Fiso})

\begin{gather}
\begin{split}&
\left.F(E)\right|_{r\rightarrow 0}=-\frac{B
\psi_0^{a/(a-2)}[(a-2)\psi_0((a-2)\psi_0+4E)+(4-4a-3a^2)(\frac{\psi_0}{\psi_0-E})^{a/(2-a)}E^2
A_4(E)]}{ 4\sqrt{2}(2-a)^2 E^{3/2}(E-\psi_0)^2 \pi^2}
\\&
\\&
A_4(E)=\ _2F_1[\frac{1}{2},\ \frac{a}{2-a},\ \frac{3}{2},\
-\frac{E}{\psi_0-E}]. \label{fe}
\end{split}
\end{gather}
We should note that this result of distribution function is only
valid in the very center of dark matter halo where the velocity
dispersion is almost isotropic. In the outer part of the halo, no
matter whether the velocity dispersion is isotropic or not, this
result is invalid and even unphysical. It's easy to note that, when
$r\rightarrow 0$, the binding energy $E$ approaches to its maximum
value $\psi_0$. Following this observation, we can further reduce
Eq.~(\ref{fe}) in the parameter space $\frac{2}{3}<a<2$ which has
covered the result of simulations. Nevertheless, it's just in this
parameter scope that F(E) in Eq.~(\ref{fe}) is physically
meaningful:
\begin{equation}
\left.F(E)\right|_{E\rightarrow
\psi_0}=C_1(\psi_0-E)^{-\frac{3}{2}-\frac{a}{2-a}},\ \frac{2}{3}<a<2
 \label{limitfe}
\end{equation}

By using equations (\ref{sigma}), (\ref{rhopsi}) and
(\ref{limitfe}), we can get the total velocity dispersion in the
very center of the halo,
\begin{gather}
\begin{split}&
\left.\sigma^2(\psi)\right|_{\psi\rightarrow
\psi_0}=\frac{2^{9/2}C_1 \pi
\psi^{5/2}}{5B}(\psi_0-\psi)^{\frac{a}{2-a}-\frac{6-a}{4-2a}}\
_2F_1[\frac{5}{2},\ \frac{1}{2}-\frac{2}{a-2},\ \frac{7}{2},\
\frac{\psi}{\psi-\psi_0}],
 \label{limitsigma}
\\&
\\&
\begin{array}{rcl}
\left._2F_1[\frac{5}{2},\ \frac{1}{2}-\frac{2}{a-2},\ \frac{7}{2},\
\frac{\psi}{\psi-\psi_0}]\right|_{\psi\rightarrow
\psi_0}&=&\frac{5(a-2)}{4(a-1)}(\frac{\psi}{\psi_0-\psi})^{-\frac{1}{2}+\frac{2}{a-2}}
\\
\\
&&
+\frac{15\sqrt{\pi}\Gamma(-2-\frac{2}{a-2})}{8\Gamma(\frac{1}{2}-\frac{2}{a-2})}(\frac{\psi}{\psi_0-\psi})^{-5/2}.
\end{array}
\end{split}
\end{gather}
Then, it's not difficult to simplify equation (\ref{limitsigma}) to
the final result:
\begin{equation}
\left.\sigma^2(\psi)\right|_{\psi\rightarrow
\psi_0}\propto\left\{
\begin{array}{cc}
(\psi_0-\psi),\ \ \ \ 2>a\geq1, \\
\\
(\psi_0-\psi)^{\frac{a}{2-a}}, \ \ \ 1>a>\frac{2}{3}.
\end{array}
\right.
 \label{sigma2}
\end{equation}
and substitute equation (\ref{limitpsir}) into equation
(\ref{sigma2}) to transform $\sigma(\psi)$ into $\sigma(r)$
\begin{equation}
\left.\sigma^2(r)\right|_{r\rightarrow 0}\propto\left\{
\begin{array}{cc}
r^{2-a},\ \ \ \ 2>a\geq1,
\\
\\
r^{a},\ \ \ \ 1>a>\frac{2}{3}.
\end{array}
\right.
 \label{sigma:r2}
\end{equation}

Now, after we know the total velocity dispersion and the density
profile in the very center of the halo, the pseudo phase-space
density can be calculated directly,

\begin{gather}
\begin{split}&
\frac{\rho(r)}{\sigma^3(r)}\propto r^{-\gamma},
\\&
\left.\gamma\right|_{r\rightarrow 0}=\left\{
\begin{array}{cc}
3-\frac{a}{2},\ \ \ \ 2>a\geq1,
\\
\\
\frac{5}{2}a,\ \ \ \ 1>a>\frac{2}{3}.
\end{array}
\right.
 \label{gamma2}
\end{split}
\end{gather}

The power-law behaviour of the pseudo phase-space density was first
found by simulation\cite{TN01} and analyzed by subsequent theoretic
work\cite{0510332,DM05,0405371,0609784}, but its physical meaning is
still unclear. For this consideration, let us compare the
distribution function and pseudo phase-space density in the very
center of the halo.

We should note that the velocity of particles, which may stably
exist in the very center of the halo, approaches to $0$ with
$r\rightarrow 0$. If the velocity term of the binding energy
Eq.~(\ref{E}) or the total velocity dispersion $\sigma^2$ is much
smaller than the second term of the RHS of Eq.~(\ref{limitpsir}) or
at most has the same asymptotic behavior as the second term of RHS
of Eq.~(\ref{limitpsir}), this condition requires $a\geq1$ which can
be obtained from Eq.~(\ref{limitpsir}) and Eq.~(\ref{sigma:r2}),
then $(\left.\psi_0-E\right)|_{r\rightarrow 0}\propto r^{2-a}$. With
this formula and Eq.~(\ref{limitfe}), we can get
\begin{equation}
\left.F(E)\right|_{r\rightarrow 0}\propto r^{-3+\frac{a}{2}}, \
a\geq1. \label{fr}
\end{equation}
From Eq.~(\ref{gamma2}) and Eq.~(\ref{fr}), we see that, if $2>a\geq
1$, the distribution function and the pseudo phase-space density
have the same asymptotic behavior where $r\rightarrow 0$. This
result indicates that the "poor-man's" phase-space density is not so
poor but might relate to real phase-space density at least in the
very center of the halo.

\section{Discussion and Conclusions}

N-body simulations show us some properties of the dark matter halo,
such as density profile $\rho(r)$, anisotropy parameter $\beta(r)$,
pseudo phase-space density, and so on. Although the simulation's
resolution is limited, these information should be considered
properly in the theoretic analysis in order to make theoretic work
more realistic.

In this paper, we investigate the problem of the center of the halo
based on the following facts: (1) the existence of universal density
profiles such as NFW\cite{NFW95,NFW96}, Moore99\cite{M99}, and in
particular the New generalized NFW profile used in our work; (2)
anisotropy parameter $\beta\approx 0$ in the center of the halo; (3)
the pseudo phase space density follows a power law in radius $r$;
(4) limited resolution in the central region.

We should notice that in this paper all results are only valid in
the very center of the halo. From a given density profile, we
calculate the asymptotic approximation of the distribution function
in the very center of the halo. Then the total velocity dispersion
and the pseudo phase-space density are obtained from the
distribution function. Eq. (\ref{gamma2}) shows the relation between
the two parameters $\gamma$ and $a$. If we set $\gamma=1.94$ as some
simulations indicate, then $a$ should be 0.776 which is smaller than
NFW's result. Otherwise, if we set $a\simeq 1\sim 1.5$ to adapt NFW
profile and Moore's profile, then $\gamma \simeq 2.25 \sim 2.5$
which is larger than the simulations' result. This apparent
contradiction between theoretic result and simulation is not so
weird since the resolution in the central region isn't high enough
to describe the very central region of the dark matter halo.

Comparing the distribution function (real phase-space density) with
the pseudo phase-space density, we find that they have the same
asymptotic behavior if $2>a\geq 1$. This result indicates that the
distribution function and pseudo phase-space density are closely
related, even though the assertion is rather premature  that the
distribution function might have a power-law behavior just as the
pseudo phase-space density in the whole range of radius $r$. This
interesting suggestion may also give us more confidence in the study
of pseudo phase-space density and some new clues to the construction
of the distribution function. By the way, some
authors\cite{0709.0434} also study the pseudo phase-space density in
from $\rho/\sigma^{3n}$. It is obvious that our conclusion in this
paper is well-founded only if $n=1$. So, from this point of view,
$\rho/\sigma^3$ may be the best choice of the pseudo phase-space
density.

For the intensive study of the dark matter halo, the full range of
radius $r$ should be involved. We need to consider how to extend the
relationship between "real" and "pseudo" phase-space density from
the very central region to other regions where the velocity
dispersion is anisotropic. Furthermore, reasonable knowledge and new
strategies should be introduced and developed for the construction
of more realistic models. First, it is admitted that dark matter
halos in the real world are not spherically symmetric. Second, the
physics in the center of the halo may be very complicated. Third,
the halo actually is a polycomponent system which contains the dark
matter, baryonic matter and even a supermassive black hole in the
center. It is not radical to say that new methods and ideas are
still needed in the future research on this matter.

\section*{Acknowledgments}
We are grateful for an anonymous referee for his/her helpful and
constructive comments to improve the manuscript. This work is
supported by the Scientific Research Foundation for the Returned
Overseas Chinese Scholars, State Education Ministry of China, and by
the Chinese Academy of Sciences under Grant No. KJCX3-SYW-N2.


\end{document}